\documentstyle[preprint,aps,epsf]{revtex}
\draft
\preprint{\bf Phys. Rev. C (1998) in press.}
\begin{document}
\title{Probing the softest region of nuclear equation-of-state}
\bigskip
\author{Bao-An Li\footnote{Address after Aug. 10: 
Dept. of Chemistry and Physics, Arkansas State University, 
P.O. Box 419, State University, AR 72467-0419} and C. M. Ko}
\address{Cyclotron Institute and Physics Department,\\
Texas A\&M University, College Station, TX 77843}
\maketitle

\begin{quote}
An attractive, energy-dependent mean-field potential for baryons is
introduced in order to generate a soft region in the nuclear
equation-of-state, as suggested by recent lattice QCD calculations
of baryon free matter at finite temperature. Based on a hadronic
transport model, we find that although this equation-of-state has
negligible effects on the inclusive hadronic spectra, it leads to a
minimum in the energy dependence of the transverse collective flow
and a delayed expansion of the compressed matter. In particular,
the transverse flow changes its direction as the colliding system
passes through the softest region in the equation-of-state.
\\
{\bf PACS number(s): 25.75.+r}
\end{quote}
\newpage
The primary goal of experiments on relativistic heavy-ion
collisions is to create and study the Quark-Gluon-Plasma (QGP)
\cite{wong}. Lattice QCD calculations of baryon-free matter at
finite temperature \cite{qcd} have shown that although both the
energy and entropy densities increase appreciably at about 150 MeV
where the phase transition from the hadronic matter to the QGP
occurs, the change in pressure is considerably weaker, leading thus
to a small sound velocity which is given by the pressure gradient
with respect to energy density, i.e., $V_s^2\equiv dP/de$. This
soft region in the nuclear equation-of-state is expected to have
significant influence on the collective dynamics of the hot and
dense matter formed in relativistic heavy ion collisions. 
In particular, a small sound velocity delays the expansion of the
compressed matter and leads also to a reduced transverse collective
flow. Indeed, recent studies based on the hydrodynamical model have
demonstrated that as one varies the incident energy of a heavy ion
collision, the system may reach the softest region in the
equation-of-state and forms a long-lived fireball, which then
results in the appearance of a minimum in the energy dependence of
the transverse collective flow \cite{bra94,hung95,ris95,plo95}. Although
the hydrodynamical model provides a convenient framework to study
the effects of nuclear equation-of-state on the collective behavior
of colliding nuclei \cite{ris95,ris96}, it requires the introduction 
of an ad hoc freeze-out procedure to compare the calculated results
with the experimental measurements.

In this Rapid Communication, we take a different approach to explore 
the effect of a softened equation-of-state on experimental observables by
using the hadronic transport model ART \cite{art}, which treats
consistently the final freeze out. The possibility of describing
the main features of lattice QCD phase transition in a hadronic
model has recently been demonstrated by allowing hadron masses to
decrease in medium, which has been argued to be related to the
restoration of chiral symmetry \cite{koch93}. Although neither QGP
phase transition nor chiral symmetry restoration actually happens
in our approach as in the hydrodynamical models, results from our
studies are useful for understanding the signals for a real phase
transition. Indeed, we find that the softened equation-of-state
leads to both a minimum in the incident energy dependence of the
transverse collective flow and a delayed expansion of the
compressed matter as in the hydrodynamical model. Our results
further show that such an equation-of-state has almost no effect on
the inclusive hadron spectra.

To introduce a soft region in the equation-of-state, we consider
for the baryon sector of the reaction system the following relation
between the pressure $P$ and energy density $e$ in the mean-field
approach\cite{clar,gale}:
\begin{eqnarray}\label{pw}
P & = & \frac{2}{3} E_k \rho+\rho \frac{dW}{d \rho }-W,\\
e & = & (m+E_k) \rho +W,
\end{eqnarray}
where $E_k$, $W$, $\rho$, and $m$ are the average kinetic energy
per baryon, potential energy density, baryon density, and nucleon
mass, respectively. It was shown in ref.\ \cite{gale} that $W$ generally
depends on both $\rho$ and $E_k$. We require that at low energy densities, 
it coincides with the normal soft equation-of-state obtained from the
well-known Skyrme interaction, i.e.,
\begin{equation}\label{hadron}
P=\frac{2}{5}E_f\rho-179\rho^2/\rho_0+164(\frac{\rho}
{\rho_0})^{7/6}\rho~~~~({\rm MeV/fm^3}),
\end{equation}
where $E_f$ and $\rho_0$ are, respectively, the Fermi energy and
normal nuclear matter density. Above the soft region, the pressure
is taken to be
\begin{equation}\label{qgp}
P=200 (\frac{\rho}{\rho_0})^{1/3}\rho-153~~~~({\rm MeV/fm^3}),
\end{equation}
which is similar to that of a noninteracting quark gas confined in
a MIT bag.

To satisfy the condition at high energy densities, the following
potential energy density is introduced,
\begin{equation}
W=\rho[600(\frac{\rho}{\rho_0})^{1/3}
-\frac{3}{5}E_f(\frac{\rho}{\rho_0})^{2/3}+153/\rho+C_1],
~~~~({\rm MeV/fm^3})
\end{equation}
where $C_1$ is a constant depending on the energy density at which
Eq. (\ref{qgp}) is reached, i.e., the upper boundary of the soft
region.

To illustrate the effects of the soft region of nuclear
equation-of-state, we consider the extreme case of a vanishing
sound velocity. This can be achieved by keeping a constant pressure
at $P_c$ while changing the energy density via the following
potential energy density in Eq. (\ref{pw}):
\begin{equation}
W=\rho(C_2-\frac{P_c}{\rho}-E_k),~~~~({\rm MeV/fm^3})
\end{equation}
where $C_2$ is again a constant determined from the energy density
of the lower boundary of transition region. In deriving the above
expression we have assumed that $E_k$ scales with $\rho^{2/3}$ as in
an adiabatic process. 

\begin{figure}[htp]
\setlength{\epsfxsize=10truecm}
\centerline{\epsffile{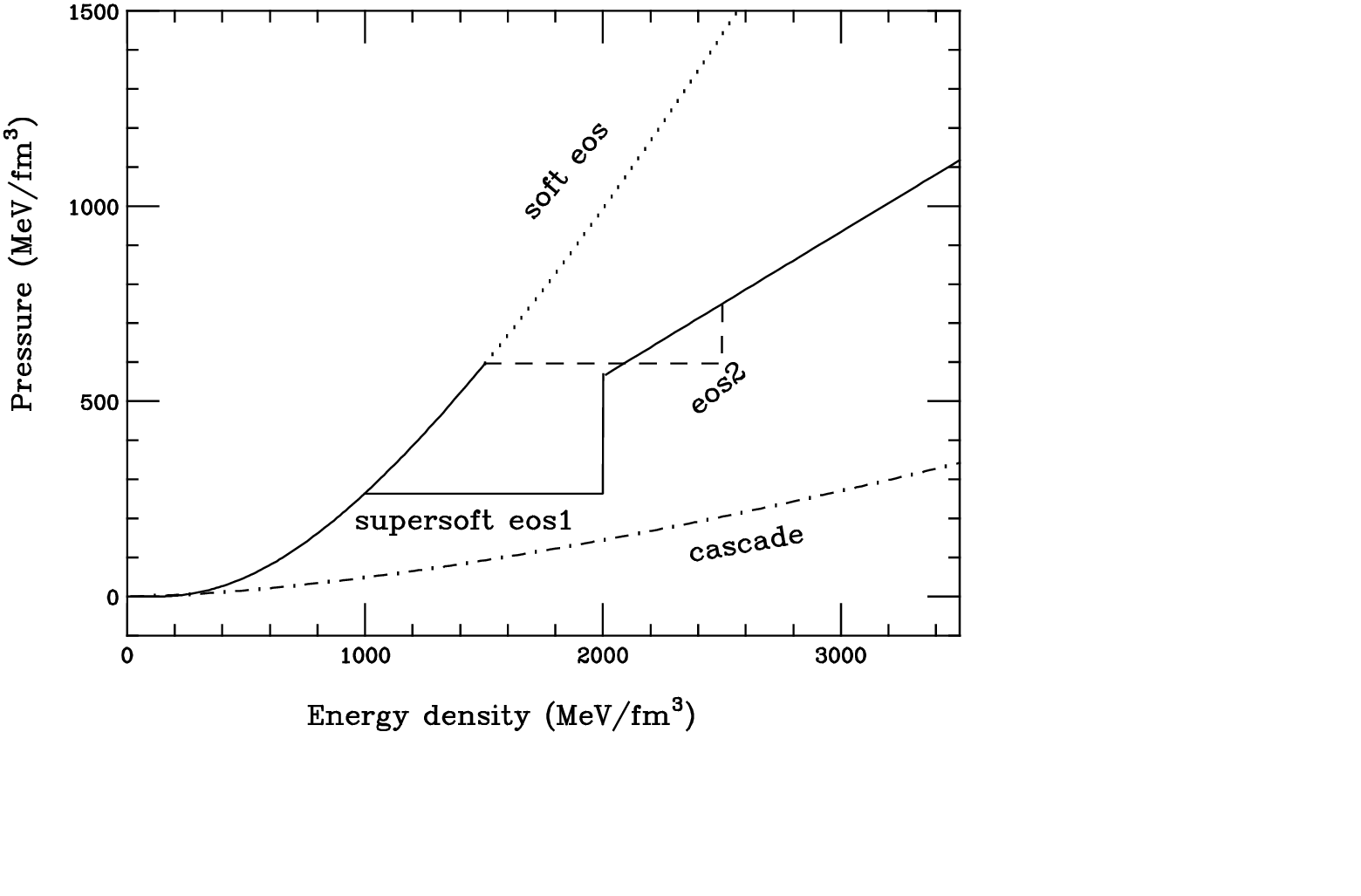}}
\caption{Nuclear equation-of-state at zero temperature.}
\label{eos} 
\end{figure}

The resulting equation-of-state is shown in Fig. \ref{eos} as
super-soft eos1 and eos2, with the softest region in the energy
density range of $1 < e < 2 ~{\rm GeV/fm}^3$ and $1.5 <e< 2.5~{\rm
GeV/fm}^3$, respectively. For comparison, we have also included the
kinetic pressure, corresponding to a pure cascade model, as a
function of energy density. The mean-field potential to be used in 
the transport model can be obtained from the relation 
$U\equiv \partial W/\partial
\rho$, which is repulsive outside but attractive inside the softest
region. In particular, the mean-field potential in the softest
region is energy-dependent in order to balance the increasing
kinetic pressure as the energy density increases. We note that both
parameters $C_1$ and $C_2$ have no effect on the dynamics as only
the second derivative of the potential energy density appears in
the equation of motion.

Effects of the softened equation-of-state can be studied using the
relativistic transport model ART. We refer the reader to Ref.\
\cite{art} for the detail of the model and its applications in
studying various aspects of relativistic heavy-ion collisions at
AGS energies. It has been found in these studies that the
transverse collective flow, defined by
\begin{equation}
<P_x/N>_t=\frac{1}{N}\int<P_x/N>(y)\cdot\frac{dN}{dy}\cdot {\rm
sgn}(y)\cdot dy,
\end{equation}
is sensitive to the nuclear equation-of-state. We expect that the
super-soft equation-of-state we introduced here will also affect
the transverse collective flow if the energy density achieved in
the collisions reaches the softest region, as in the recent study
based on the hydrodynamical model \cite{ris95,ris96}. In the latter
study, a minimum in the excitation function of $<P_x/N>_t$ at a
beam energy of about 6 GeV/nucleon has been found when a softened
equation-of-state due to the QCD phase transition is used.

\begin{figure}[htp]
\setlength{\epsfxsize=10truecm}
\centerline{\epsffile{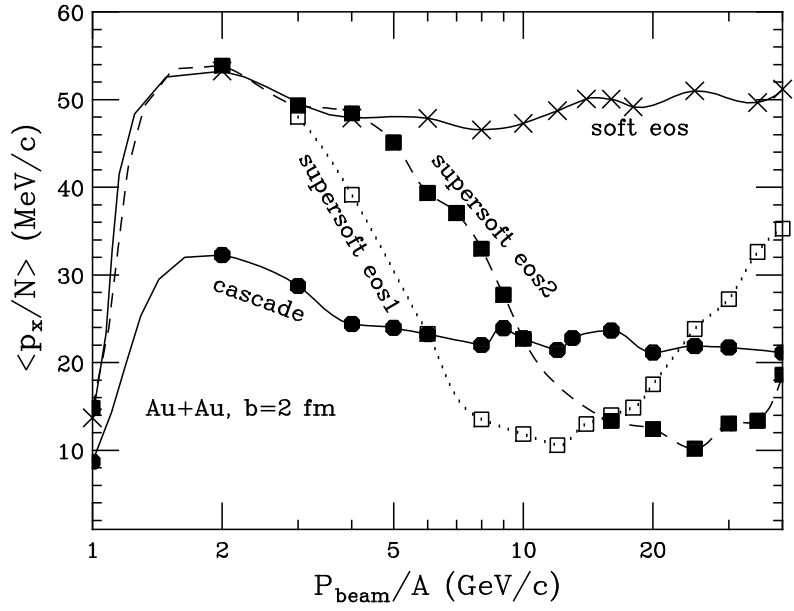}}
\caption{Excitation function of total in-plane transverse momentum
for the reaction of Au+Au at an impact parameter of 2 fm.}
\label{exc} 
\end{figure}

\begin{figure}[htp]
\setlength{\epsfxsize=10truecm}
\centerline{\epsffile{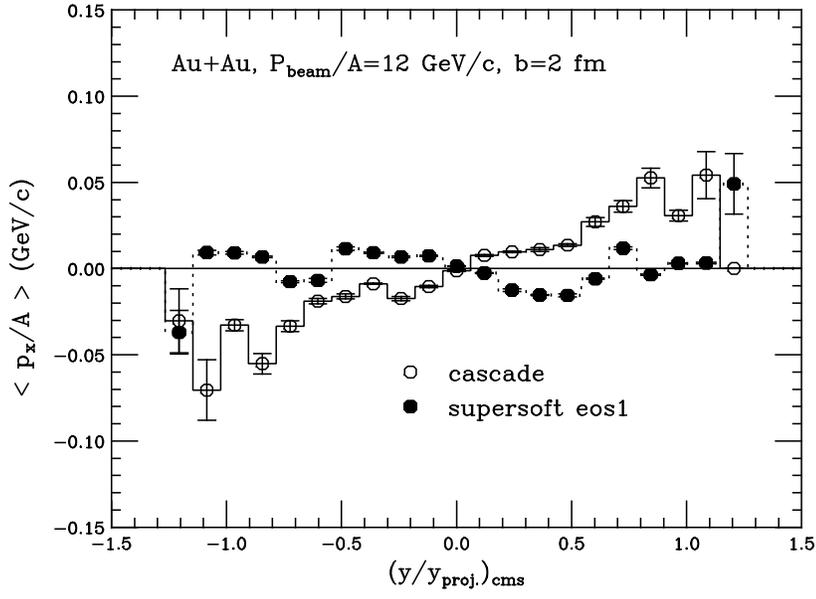}}
\caption{In-plane transverse momentum distribution
for the reaction of Au+Au at $P_{\rm beam}/A=12$ GeV/c and an
impact parameter of 2 fm.}
\label{flow} 
\end{figure}

In Fig. \ref{exc}, we show the excitation function for the average
transverse momentum predicted by the ART model using the four
equation-of-states shown in Fig.\ 1. For the case of either the
cascade or the normal soft equation-of-state, details can be found
in our previous publications \cite{art}. It is interesting to see
that the super-soft equation-of-state with the softest region leads
to a minimum in the excitation function of the transverse flow, as
in hydrodynamical calculations. Furthermore, the incident energy at
which the minimum occurs depends on the location of the softest
region in the equation-of-state. For both super-soft eos1 and eos2,
the average transverse momenta near the minima are even smaller
than those in the cascade calculations as attractive mean-field
potentials have been introduced in these equation-of-states to
counterbalance the positive gradient of the thermal pressure. This
is more clearly seen in Fig. \ref{flow}, where we compare the
average transverse momentum distribution using the cascade model
and the super-soft eos1 for the Au+Au reaction at $P_{\rm
beam}/A=12$ GeV/c. It is seen that the flow parameter, defined as
the slope of the transverse momentum distribution at mid-rapidity,
changes from positive to negative sign as a result of the
attractive mean-field potential in the softest region.
Our results also show that the flow parameter changes gradually
from positive to negative then to positive as the beam energy
increases. A change in the sign of flow parameter as the beam
energy increases is another clear indication of the softening of
the nuclear equation-of-state. To observe this effect
experimentally, one needs to measure the absolute direction of the
transverse flow besides its strength. This can be done by studying
the shadowing effect on pion flow \cite{gos89,bal91,bal94,dan} and the 
circular polarization of the $\gamma$ ray emitted from the reaction
\cite{betty}, as in heavy ion collisions at low energies.

\begin{figure}[htp]
\setlength{\epsfxsize=14truecm}
\centerline{\epsffile{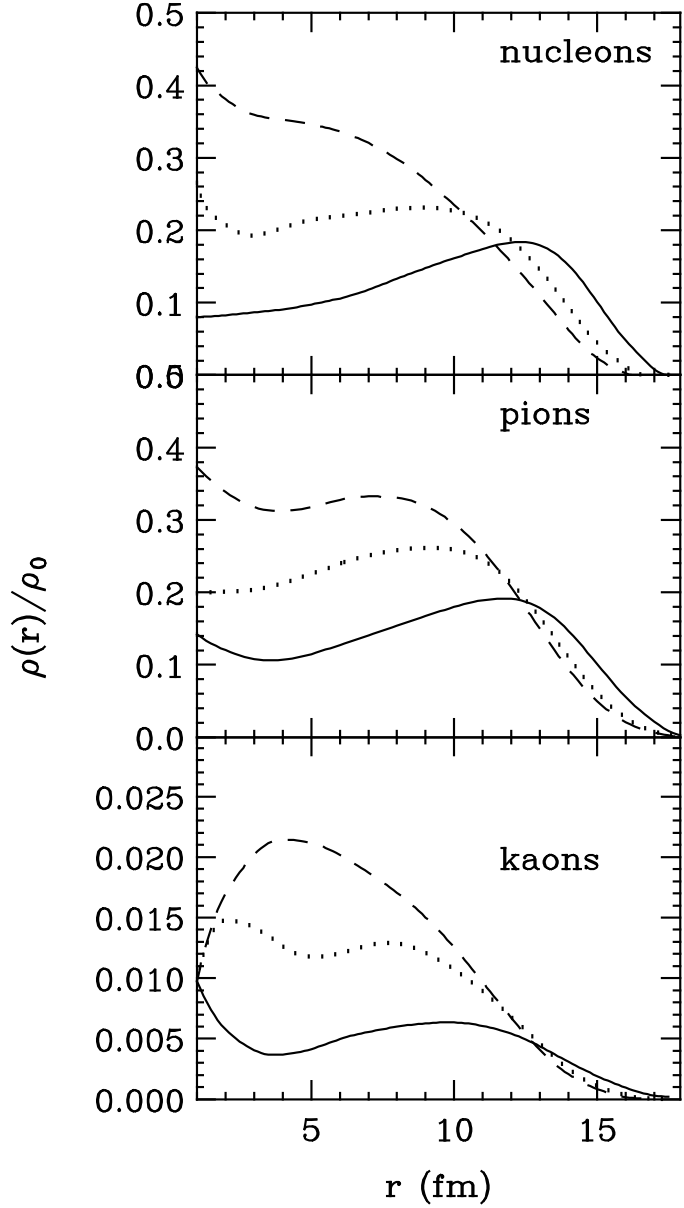}}
\caption{Radial density distribution of nucleons, pions, and kaons in the
reaction Au+Au at $P_{\rm beam}/A=12$ GeV/c and an impact parameter
of 2 fm. The solid, dot, and dash lines are calculated using the
soft eos, the super-soft eos1, and the cascade model,
respectively.}
\label{density} 
\end{figure}

A softened equation-of-state is expected not only to slow down the
expansion of the system but also to affect the amount of
compression in the collision. Since the pressure decreases 
from the normal soft, super-soft to the cascade at the same energy density,
one expects to reach the highest compression in the cascade model.
Shown in Fig. \ref{density} are the radial density distributions of
nucleons, pions and kaons at 20 fm/c from the Au+Au reaction at an
impact parameter of 2 fm and a beam momentum of 12 GeV/c. At this
already rather late time of the reaction, the normal soft
equation-of-state gives the most extend particle distributions,
indicating thus the fastest compression and expansion. The particle
distributions from calculations using the super-soft eos1 are more
compact, implying thus a slower compression and expansion.

It is worth mentioning that we have also studied other inclusive
hadronic observables using the four equations of state and have
found that it is almost impossible to distinguish the predictions
using the softened equation-of-state from others.

In summary, we have studied effects of the softened nuclear
equation-of-state on heavy ion collisions at relativistic energies
within a hadronic transport model. To keep a zero pressure gradient
in the softest region, an attractive mean field for baryons has
been introduced. The softened equation-of-state is found to lead to
a minimum in the excitation function of transverse flow and a
delayed expansion but has almost no effect on the inclusive
hadronic observables. Although it is questionable to use the
hadronic degrees of freedom at very high energy densities, our
results do indicate that a super-soft equation-of-state can produce
the effects expected from the phase transition. Experiments on the
measurement of the transverse collective flow at different incident
energies will be extremely useful in our search for the signatures
of chiral and deconfinement transitions.

This work was supported in part by NSF Grant No. PHY-9509266, the
Robert A Welch foundation under Grant A-1358, and the Texas
Advanced Research Program.

\end{document}